\documentclass[aps,prb,twocolumn,superscriptaddress,showpacs,reprint]{revtex4}
\usepackage{multirow}
\usepackage{graphicx}
\usepackage{longtable}
\usepackage[utf8]{inputenc}
\usepackage{epstopdf}
\usepackage{longtable}

\begin{document}
\title{Molecular Anisotropic Magnetoresistance}

\author{Fabian Otte}
%\email{otte@theo-physik.uni-kiel.de}
\affiliation{Institut f\"ur Theoretische Physik und Astrophysik, Christian-Albrechts-Universit\"at zu Kiel, D-24098 Kiel, Germany}
\author{Stefan Heinze}
\affiliation{Institut f\"ur Theoretische Physik und Astrophysik, Christian-Albrechts-Universit\"at zu Kiel, D-24098 Kiel, Germany}
\author{Yuriy Mokrousov}
\affiliation{Peter Gr\"unberg Institut and Institute for Advanced Simulation, Forschungszentrum J\"ulich and JARA, D-52425 J\"ulich, Germany}

\date{\today}
\begin{abstract}
 
Using density functional theory calculations, we demonstrate that the effect of anisotropic magnetoresistance (AMR) can be enhanced
by orders of magnitude with respect to conventional bulk ferromagnets in junctions containing molecules sandwiched between ferromagnetic
leads. We study ballistic transport in metal-benzene complexes contacted by $3d$ transition-metal wires. We show 
that the gigantic AMR can arise from spin-orbit coupling effects in the leads, drastically enhanced by orbital-symmetry filtering properties of the molecules.
We further discuss how this molecular anisotropic magnetoresistance (MAMR) can be tuned by proper choice of materials and their electronic properties. 
 
\end{abstract}

\maketitle

Anisotropic magnetoresistance (AMR) in ferromagnets was discovered 
two centuries ago by Thomson~\citep{thomson}, but
it took more than a hundred years to exploit this phenomenon to full extent and to provide a theoretical understanding~\citep{smit, potter, mcg_potter, fert}. In the past decades, this relativistic effect, the essence of which lies in
the dependence of the longitudinal conductance on the
direction of the magnetization in a ferromagnet or staggered magnetization in an antiferromagnet due to spin-orbit coupling (SOC) has become one of the main tools used to access and probe the magnetization
dynamics in various spintronics setups~\cite{PhysRevLett.107.086603,2015arXiv150303765W}. 

AMR in bulk ferromagents has been extensively explored 
in the past~\cite{mcg_potter}. In tunneling
geometry the AMR has been measured in ultrathin films utilizing the scanning tunneling microscope~\citep{PhysRevLett.89.237205} and in planar tunnel junctions with
ferromagnetic semiconductors~\citep{PhysRevLett.93.117203}. The phenomenon of ballistic AMR has been theoretically suggested to exist in the contact limit of nano-scale atomic junctions~\citep{PhysRevLett.94.127203}, 
and it was eventually demonstrated experimentally in Co break-junctions~\citep{Sokolov}. 

AMR of single-atom and single-molecule contacts
is caused by the magnetization-direction sensitive, SOC-induced hybridization between specific orbitals at a certain energy. 
AMR is severely limited by the states which are not altered by the change of magnetization direction and
thus provide a constant background conductance.
In typical bulk ferromagnets, it constitutes only a few percent.
In the ballistic regime, suppressing the transmission due to SOC-insensitive states and promoting the conductance due to states which are
strongly dependent on the magnetization direction, 
promises enhanced values of AMR.

This scenario 
can be realized with molecules sandwiched between magnetic leads since molecules often bind rather weakly to 
the substrates on which they are deposited and their orbitals preserve this 
initial symmetry and localized molecular orbital character~\citep{PhysRevLett.113.106102,rectifier}.
This makes them ideally suited for filtering out states of a specific symmetry and  
favoring the transmission due to lead states heavily influenced by SOC.
While molecules are well-known for their spin-filtering properties~\citep{PhysRevLett.97.097201} and AMR of molecular systems in the tunneling regime 
has been investigated 
experimentally~\citep{PhysRevB.84.125208,Jijun},  magnetic anisotropy effects in organic materials are usually believed to be small due to weak spin-orbit interaction
of light elements and are scarcely explored~\citep{C3CC47126H}. 

\begin{figure}
\includegraphics[width=\linewidth]{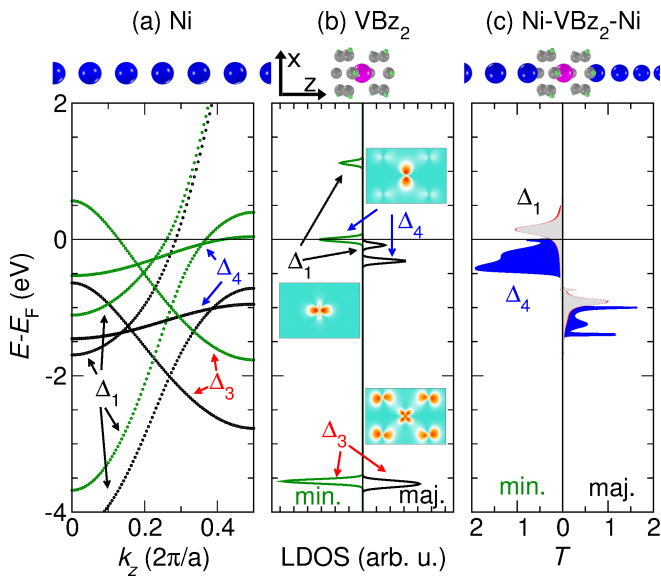}
\caption[fig1]{(a) Bandstructure of a Ni monowire. Majority and minority states are marked with black and green circles, respectively.
(b) Spin-decomposed local density of states (LDOS) of VBz$_2{}$ molecule. The insets from top to bottom show the charge density plots of the molecular orbitals in the $xz$-plane.
(c) Spin-decomposed transmission $T$ of VBz$_2{}$ molecule contacted by Ni monowires. The contributions of orbitals of $\Delta_1{}$, $\Delta_3{}$ and $\Delta_4{}$ symmetry are colored in gray, red and blue, respectively. 
SOC has been neglected.} 
\label{fig:scheme}
\end{figure}

In this work, by density functional theory (DFT) calculations, we demonstrate that AMR in the ballistic regime can be enhanced by orders of magnitude 
via combining SOC-triggered modifications in the electronic structure of transition-metal wires with the orbital-symmetry filtering properties
of molecules. This leads to the effect of {\it molecular anisotropic magnetoresistance} (MAMR), defined as:
\begin{equation}
\label{eq:MAMR}
\mathrm{MAMR}= \frac{T_{\parallel} - T_{\perp}} {T_{\perp}},
\end{equation}
where $T_{\perp}{}$ and $T_{\parallel}{}$ are the transmissions for perpendicular and parallel orientation of the magnetization with respect to the
chain axis, respectively. To demonstrate the effect, we focus on metal-benzene complexes of the VBz$_2$-type 
(Bz, VBz$_2{}$, V$_2$Bz$_3{}$, TaBz$_2{}$, and NbBz$_2{}$), 
some of which have been studied before both theoretically and experimentally~\citep{541378242,0957-4484-18-49-495402}, sandwiched between 
leads which are modeled by Ni and Co linear monoatomic chains (see Fig.~\ref{fig:scheme}~(c) for a sketch of the geometry). 

The method of study~\citep{PhysRevB.85.245412,PhysRevB.86.165449} is based on density-functional theory in the local density approximation~\citep{doi:10.1139/p80-159},
 and utilizes the one-dimensional version~\citep{PhysRevB.72.045402} of the full-potential linearized 
augmented plane-wave method as implemented in \texttt{FLEUR}~\footnote{www.flapw.de}. 
The leads were described in a single-atom unit cell with a lattice constant of 2.21\,\AA.
The scattering region was modeled by a symmetric unit cell consisting of up to ten Ni (Co) atoms and the molecule.
For the VBz$_2{}{}$ molecule the distance was set to the relaxed values of the isolated molecule~\citep{0957-4484-18-49-495402}.
The same parameters were used for the other molecules since they only change by a few percent when the central metal atom is replaced~\citep{0957-4484-18-49-495402}.

We begin the demonstration of the concept of MAMR by analyzing the electronic bandstructure of an isolated Ni monowire (MW) and the local density of states (LDOS) of the VBz$_2{}$ molecule $-$ systems, which we intend to bring in contact $-$ presented  in Fig.~\ref{fig:scheme}~(a) and (b) neglecting the effect of spin-orbit coupling. 
The exchange splitting of the minority and majority Ni states, visible in~Fig.~\ref{fig:scheme}~(a), leads to a magnetic moment of about 1.1\,$\mu_{\mathrm{B}}{}$ 
of the Ni atoms. The critical bands of the Ni chain in the following discussions will be the highly dispersive bands of $\Delta_1{}$ ($s, d_{z^2}$) symmetry and of the very localized $\Delta_4{}$ ($d_{xy}, d_{x^2-y^2}$) symmetry. 
The bands of $\Delta_3{}$ ($d_{xz}, d_{yz}$) symmetry are 
of minor importance since the molecular orbitals (MO) of VBz$_2{}$ of $\Delta_3{}$-symmetry are very low in energy. In contrast, the VBz$_2{}$ MOs of  $\Delta_1{}$- and $\Delta_4{}$-symmetry lie close to
the Fermi level and are thus able to interact with Ni MW states of the corresponding symmetry. In the isolated molecule, see~Fig.~\ref{fig:scheme}~(b), the splitting of minority and majority states results in a V magnetic moment of 1.0\,$\mu_{\mathrm{B}}{}$.

In Fig.~\ref{fig:scheme}~(c) we show the resulting spin-resolved transmission of VBz$_2{}$ contacted by Ni MWs with 
a distance between the Ni and V atom of 
$d_{\mathrm{Ni-V}}=4.232$\,\AA\, on both sides. At this distance the largest value of MAMR due to optimal bonding characteristics
is found as will be discussed at the end of this paper. 
In realistic systems, it has been shown that the bonding characteristics can be tuned by graphene spacer layers~\citep{rectifier} or the chemical structure of the ligands of
a magnetic complex~\citep{PhysRevLett.113.106102}.
The transmission exhibits localized contributions of $\Delta_1{}$-symmetry in the minority and majority channel at +0.13 eV
and $-0.93$ eV, respectively, nearly preserving the exchange splitting between majority and minority $\Delta_1{}$-MOs  of +1.4 eV in the free molecule displayed in 
Fig.~\ref{fig:scheme}~(b).
The states of $\Delta_4{}$-symmetry, on the other hand, contribute by broad, band-like features from $-0.5{}$\,eV to $E_F$ and from $-1.4{}$ to $-1.0{}$ eV in the minority and majority channels, respectively,
coinciding precisely with the regions of $\Delta_4{}$ Ni bands in the leads.
This shape of the transmission can be understood from the observation that the $\Delta_4{}$-states of the Ni MW hybridize strongly with the
respective MOs of VBz$_2{}$, while the
hybridization of the states of $\Delta_1{}$ and $\Delta_3{}$ symmetry is very small. This is caused by the already mentioned large energy offset between the $\Delta_3$-states of the transition metal wire and the molecule and, in addition, the shape of the
charge densities of the MOs in the $xz$-plane displayed as 
insets in Fig.~\ref{fig:scheme}(b). While the charge density of the MO of $\Delta_1$-symmetry is only localized around the
V atom, the one of the MO of $\Delta_4$-symmetry shows considerable contributions at the benzene ring thus 
reducing the effective bonding distance for the $\Delta_4$-states compared to the $\Delta_1$-states. The strong effect of the symmetry-dependent hybridization between the lead states and the MOs of the molecule thus not only 
leads to the effect of spin-filtering as pointed out before~\citep{PhysRevLett.97.097201}, but also gives rise to the orbitally-filtered transmission. 

\begin{figure*}
\includegraphics[width=0.8\linewidth]{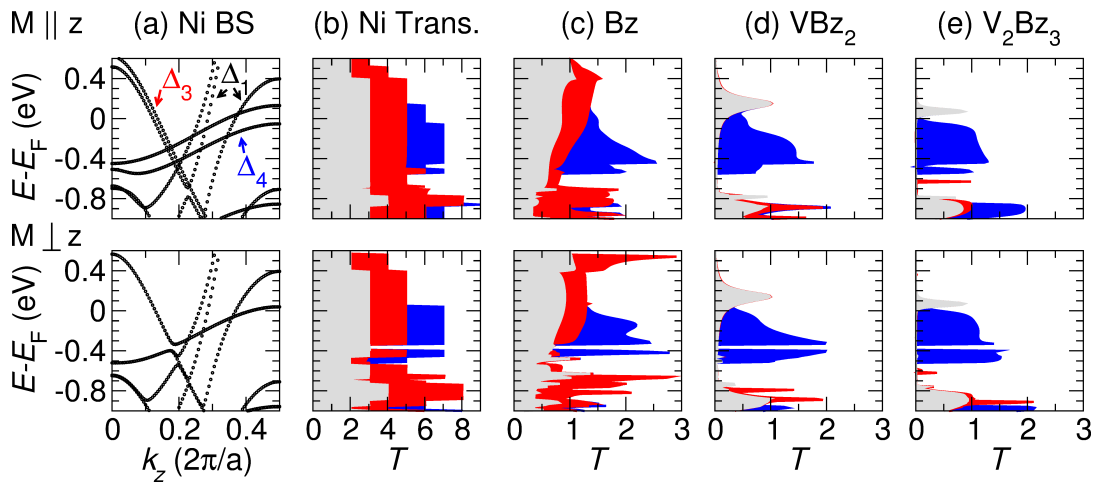}
\caption[fig2]{(a) Ni chain bandstructure with SOC for $\mathbf{M}\parallel z{}$ (top) and $\mathbf{M} \perp z{}$ (bottom) where $z$ denotes the chain axis.
   (b) Transmission of an infinite Ni wire for $\mathbf{M}\parallel z{}$ (top) and $\mathbf{M} \perp z{}$ (bottom). 
     (c) Transmission of Bz molecule contacted by Ni monowires for $\mathbf{M}\parallel z{}$ (top) and $\mathbf{M} \perp z{}$ (bottom). 
     (d) shows the same as (c) of VBz$_2{}$ molecule.
     (e) shows the same as (c) of V$_2$Bz$_3$ molecule. The color code is as in Fig.~\ref{fig:scheme}.}
\label{fig:Bz_VBz_V2Bz3}
\end{figure*}
Considering spin-orbit coupling, the modifications in the electronic structure of the Ni chain
upon changing the magnetization from along-the-chain axis ($\mathbf{M}\parallel z{}$) to perpendicular ($\mathbf{M}\perp z{}$) are clearly
visible in Fig.~\ref{fig:Bz_VBz_V2Bz3}~(a). 
We focus on an energy region which leads to the largest MAMR.
In particular, at the point where the $\Delta_4{}$- and 
$\Delta_3{}$-bands cross at $-0.38{}$\,eV we observe an avoided degeneracy as the magnetization
is rotated from along-the-chain axis to perpendicular, since for $\mathbf{M}\perp z{}$ the $\Delta_3{}$- and $\Delta_4{}$-states 
are allowed to couple due to non-zero matrix elements of spin-orbit coupling between them for this
magnetization direction~\citep{PhysRev.140.A1303}. 
As a result, were the Fermi energy 
located at $-0.38{}$\,eV, the ballistic AMR of an infinite Ni chain of 130\,\% would be significant, but its value would 
be hindered by the unavoidable large contribution to the transmission from magnetization-direction insensitive 
$\Delta_1{}$-states at this energy (cf.~Fig.~\ref{fig:Bz_VBz_V2Bz3}~(a)). 

The situation is completely different for our Ni-VBz$_2{}$-Ni system.
The magnetization-direction induced changes in the bandstructure of the Ni chain are directly passed on to the full transmission of the junction, see Fig.~\ref{fig:Bz_VBz_V2Bz3}~(d). For $\mathbf{M}\perp z{}$ the 
$\Delta_4{}$-transmission is quenched between $-0.4{}$ and $-0.33{}$\,eV, however, in contrast to the case of the infinite Ni chain (cf.~Fig.~\ref{fig:Bz_VBz_V2Bz3}~(b)), the remaining transmission of the order of 0.05 in this energy interval is due to the tail of localized $\Delta_1{}$-originated transmission peaks at 0.1 and $-0.9{}$\,eV. 
For $\mathbf{M}\parallel z{}$, on the other hand, the transmission at this energy has a value of about 1.5\,, originating mainly in $\Delta_4{}$-states. The gigantic magnitude of the AMR for this junction of the 
order of $3.5\times 10^3$\,\% nicely illustrates the principle of employing the symmetry-filtering properties of the 
molecules for arriving at {\it molecular} AMR. 

The strength of the filtering effect and MAMR can be controlled by the size of the contacted molecule. 
For a single benzene molecule contacted by Ni MWs, Fig.~\ref{fig:Bz_VBz_V2Bz3}\,(c)~\footnote{The distance between the
 apex Ni atom and Bz is 2.428\,\AA, the same distance as in the Ni-VBz$_2{}$-Ni junction with $d_{\mathrm{Ni-V}}=4.232$ \AA.}, the transmission due to $\Delta_1{}$- and $\Delta_3{}$-states displays
band-like features in a wide energy interval due to strong hybridization of the Ni chain with Bz, and of Ni apex atoms across benzene. This leads to poor filtering of $\Delta_1{}$- and $\Delta_3{}$-states, leaving a transmission
of about 0.7\, for $\mathbf{M}\perp z{}$ in the energy range from $-0.4{}$ to $-0.33{}$\,eV, while the 
transmission for $\mathbf{M} \parallel z{}$ is about 2.2\,. This results in MAMR of about 230\,\%, which is one order of magnitude smaller as compared to the Ni-VBz$_2{}$-Ni junction. 

On the other hand,
the filtering effect can be enhanced by considering a longer V$_2$Bz$_3{}$ molecule instead of VBz$_2{}$
(while keeping $d_{\mathrm{Ni-V}}$ the same). For this system the $\Delta_1{}$ transmission peak at +0.05 eV in  Fig.~\ref{fig:Bz_VBz_V2Bz3}~(e) becomes very sharp, while the energetic tails of both $\Delta_1$-peaks at +0.05 and $-0.9{}$\,eV are strongly suppressed. As a result, the transmission between $-0.4{}$ and $-0.33{}$\,eV is almost solely determined by the states of $\Delta_4{}$-symmetry. The total transmission in this energy region for 
$\mathbf{M}\perp z{}$ is only of the order of $10^{-3}{}$\,, while it constitutes 1.2\, for $\mathbf{M}\parallel z{}$, which results in MAMR of about $1.3\times 10^5{}$\,\% $-$ two orders of magnitude larger as compared to the VBz$_2{}$ junction.

\begin{figure}
\includegraphics[width=\linewidth]{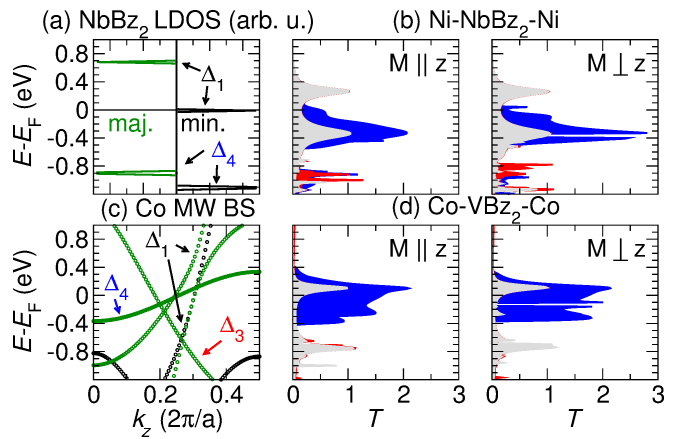}
\caption[fig3]{(a) Spin-decomposed local density of states of NbBz$_2{}$ molecule without SOC. 
 (b) Transmission of NbBz$_2{}$ molecule contacted by Ni monowires with SOC for $\mathbf{M}\parallel z{}$ (left) and $\mathbf{M} \perp z{}$ (right). 
 (c) Bandstructure of a Co monowire without SOC. 
 (d) shows the same as (b) of VBz$_2{}$ molecule contacted by Co monowires. The color code is as in Fig.~\ref{fig:scheme}. } 
\label{fig:Nb_Ta_Co}
\end{figure}
The quality of the filtering properties of such molecular junctions expectedly depends very strongly on the energetic position of the MOs of the contacted molecule with respect to the lead states. 
We demonstrate this by using the example of a Ni-NbBz$_2{}$-Ni junction. As apparent from Fig.~\ref{fig:Nb_Ta_Co}~(a), 
the LDOS of the isolated NbBz$_2{}$ is very similar to that of the free VBz$_2{}$ since Nb is isolectronic to V. 
However, the exchange splitting of $\Delta_1$-orbitals is only half as large for NbBz$_2{}$, which is 
reflected in the reduced value of the magnetic moment of 0.67\,$\mu_{\mathrm{B}}{}$. 
As in the case of VBz$_2{}$, the exchange splitting between the $\Delta_1$-states is nearly conserved when the molecule is contacted by Ni leads, Fig.~\ref{fig:Nb_Ta_Co}~(b). 
Therefore, the majority $\Delta_1$-peak in transmission lies at $-0.36{}$ eV $-$ directly in the energy region of the avoided level crossing in the bandstructure of the Ni chain discussed above.
This severely hinders the orbital filtering in this energy region, resulting in MAMR of only 90\,\% despite 
a significantly stronger spin-orbit interaction of Nb atoms, as compared to Ni or V. 

A similar value for the MAMR of 100\,\% is obtained by replacing Nb with Ta for which SOC is even more enhanced. 
This can be explained by the fact that within the scenario for MAMR that we consider here, it is the SOC-induced modifications in the electronic structure of the leads which are promoted by the molecule. 
This is also visible when the Ni leads are replaced by chains consisting of Co atoms. In the Co MW bandstructure (Fig.~\ref{fig:Nb_Ta_Co}~(c)) the bands of $\Delta_3{}$- and $\Delta_4{}$-symmetry cross at 
$-0.13{}$\,eV for the case without SOC. This degeneracy is lifted at that energy for $\mathbf{M} \perp z{}$ when SOC is included, in analogy to the Ni case. Since Co is slightly lighter than Ni the energy gap devoid of $\Delta_3{}$- and $\Delta_4{}$-states for  $\mathbf{M}\perp z{}$ 
occurs from $-0.1{}$ eV to $-0.15{}$ eV,\,i.e., it  
extends over a range of 50 meV, as compared to 67 meV for Ni. The MAMR in this energy region is of about 1800\,\%, of a similar magnitude as for the same molecule contacted by Ni, Fig.~\ref{fig:Nb_Ta_Co}~(d).

So far, we discussed the energy region providing the largest effect for the considered molecular junctions.
However, an enhanced AMR is also observed at other energies. For example, the AMR at the Fermi energy, which is due to splitting 
of the $\Delta_4$ bands and amounts to $-18$~\% for the infinite Ni MW, increases to $+30$\% for the Ni-VBz$_2$-Ni junction 
and $-85$~\% for Ni-V$_2$Bz$_3$-Ni and to $+67$\% for Ni-NbBz$_2$-Ni (cf.~Figs.~\ref{fig:Bz_VBz_V2Bz3} and \ref{fig:Nb_Ta_Co}). 
The sign change for the cases of VBz$_2$ and NbBz$_2$ illustrates the sensitivity of the AMR on the details of the alignment 
of molecular orbitals with lead states. 
However, the key effect of the orbital filtering and respective $\Delta_1$-background contribution reduction is also responsible for the enhanced AMR at the Fermi energy.

\begin{table}
\caption{\label{tab1} MAMR at $-0.36{}$ eV of a VBz$_2{}$ molecule sandwiched between Ni leads with the distance $d_{\mathrm{Ni-V}}{}$ ranging from 
2.96 to 6.25\,\AA. }.
\begin{ruledtabular}
\begin{tabular}{lccccc}
$d_{\mathrm{Ni-V}}$ (\AA)           & 2.96 & 3.70 & 4.23  & 4.76  & 6.35 \\
MAMR (\%)  &  0   & 2500    &  3500    & 74    &   3    \\     
\end{tabular}
\end{ruledtabular}
\end{table}
In the examples considered above the states to be suppressed were of $\Delta_1$-symmetry. However, if our aim was 
to promote the crossing of majority $\Delta_1$- and minority $\Delta_3$-states at $-$0.7\,eV in the electronic structure of Ni (see 
Fig.~\ref{fig:Bz_VBz_V2Bz3} (a) at $k_z\approx 0.22 \times 2 \pi/a$), it would be the background transmission due to majority 
$\Delta_3$- and minority $\Delta_1$-states which would have to be suppressed by orbital filtering of the 
corresponding molecule. 

One of the readily achievable experimental tools to tune the parameters which control the magnitude of the MAMR
is the distance between the leads and the molecule. In Table~\ref{tab1} we show the development of MAMR of a Ni-NbBz$_2$-Ni junction as the distance $d_{\mathrm{Ni-V}}{}$ is changed between 2.96\,\AA{} and 6.25\,\AA, while keeping
the energy at $-0.36{}$\,eV. At small distance the MAMR is negligible, while it rises to its maximal value at $d_{\mathrm{Ni-V}}{}$ between 
1.96\,\AA{} and 2.24\,\AA, and then rapidly decays as the distance is increased further.
This behavior can be attributed to the 
strength of hybridization between states of different symmetry from the Ni MW and VBz$_2{}$, 
and its variation with $d_{\mathrm{Ni-V}}{}$. At small distances the hybridization between the 
$\Delta_1{}$-, $\Delta_3{}$- and $\Delta_4{}$-states is very strong, leading to a large background transmission which quenches the $\Delta_4{}$-MAMR, in a similar way to the Ni-Bz-Ni junction.
The hybridization reduces strongly for states of $\Delta_1{}$ and $\Delta_3{}$ symmetry at intermediate distances, 
while it remains strong for the $\Delta_4{}$-states, as discussed previously, promoting thus $\Delta_4{}$-MAMR. 
Eventually, at very large  $d_{\mathrm{Ni-V}}{}$ the overlap between  $\Delta_4{}$-states also vanishes, with tunneling
across $\Delta_1{}$-states dominating the tiny MAMR.

In conclusion, our calculations demonstrate the phenomenon of huge MAMR and point out the importance of the modifications
in the electronic structure of the leads, contacted through a molecule, upon varying their magnetization direction. The
magnitude of the MAMR sensitively depends on the details of the orbital filtering by the molecule. The specific 
electronic structure of the molecule and its coupling to the leads should be engineered based on the condition as to 
which part of the electronic structure in the leads should be promoted, and which parts of it should be suppressed.

\acknowledgements{ S.H. and F.O. thank the DFG for funding within the SFB 677.}


\begin{thebibliography}{24}
\expandafter\ifx\csname natexlab\endcsname\relax\def\natexlab#1{#1}\fi
\expandafter\ifx\csname bibnamefont\endcsname\relax
  \def\bibnamefont#1{#1}\fi
\expandafter\ifx\csname bibfnamefont\endcsname\relax
  \def\bibfnamefont#1{#1}\fi
\expandafter\ifx\csname citenamefont\endcsname\relax
  \def\citenamefont#1{#1}\fi
\expandafter\ifx\csname url\endcsname\relax
  \def\url#1{\texttt{#1}}\fi
\expandafter\ifx\csname urlprefix\endcsname\relax\def\urlprefix{URL }\fi
\providecommand{\bibinfo}[2]{#2}
\providecommand{\eprint}[2][]{\url{#2}}

\bibitem[{\citenamefont{Thomson}(1856)}]{thomson}
\bibinfo{author}{\bibfnamefont{W.}~\bibnamefont{Thomson}},
  \bibinfo{journal}{Proc. R. Soc.} \textbf{\bibinfo{volume}{8}},
  \bibinfo{pages}{546} (\bibinfo{year}{1856}).

\bibitem[{\citenamefont{Smit}(1951)}]{smit}
\bibinfo{author}{\bibfnamefont{J.}~\bibnamefont{Smit}},
  \bibinfo{journal}{Physica} \textbf{\bibinfo{volume}{17}},
  \bibinfo{pages}{612} (\bibinfo{year}{1951}).

\bibitem[{\citenamefont{Potter}(1974)}]{potter}
\bibinfo{author}{\bibfnamefont{R.~I.} \bibnamefont{Potter}},
  \bibinfo{journal}{Phys. Rev. B} \textbf{\bibinfo{volume}{10}},
  \bibinfo{pages}{4626} (\bibinfo{year}{1974}).

\bibitem[{\citenamefont{McGuire and Potter}(1975)}]{mcg_potter}
\bibinfo{author}{\bibfnamefont{T.~R.} \bibnamefont{McGuire}} \bibnamefont{and}
  \bibinfo{author}{\bibfnamefont{R.~I.} \bibnamefont{Potter}},
  \bibinfo{journal}{IEEE Trans. Magn.} \textbf{\bibinfo{volume}{11}},
  \bibinfo{pages}{1018} (\bibinfo{year}{1975}).

\bibitem[{\citenamefont{Fert and Campbell}(1976)}]{fert}
\bibinfo{author}{\bibfnamefont{A.}~\bibnamefont{Fert}} \bibnamefont{and}
  \bibinfo{author}{\bibfnamefont{I.~A.} \bibnamefont{Campbell}},
  \bibinfo{journal}{J Phys. F: Met. Phys.} \textbf{\bibinfo{volume}{6}},
  \bibinfo{pages}{849} (\bibinfo{year}{1976}).

\bibitem[{\citenamefont{Seemann et~al.}(2011)\citenamefont{Seemann, Freimuth,
  Zhang, Bl\"ugel, Mokrousov, B\"urgler, and
  Schneider}}]{PhysRevLett.107.086603}
\bibinfo{author}{\bibfnamefont{K.~M.} \bibnamefont{Seemann}},
  \bibinfo{author}{\bibfnamefont{F.}~\bibnamefont{Freimuth}},
  \bibinfo{author}{\bibfnamefont{H.}~\bibnamefont{Zhang}},
  \bibinfo{author}{\bibfnamefont{S.}~\bibnamefont{Bl\"ugel}},
  \bibinfo{author}{\bibfnamefont{Y.}~\bibnamefont{Mokrousov}},
  \bibinfo{author}{\bibfnamefont{D.~E.} \bibnamefont{B\"urgler}},
  \bibnamefont{and} \bibinfo{author}{\bibfnamefont{C.~M.}
  \bibnamefont{Schneider}}, \bibinfo{journal}{Phys. Rev. Lett.}
  \textbf{\bibinfo{volume}{107}}, \bibinfo{pages}{086603}
  (\bibinfo{year}{2011}).

\bibitem[{\citenamefont{{Wadley} et~al.}(2015)\citenamefont{{Wadley},
  {Howells}, {Zelezny}, {Andrews}, {Hills}, {Campion}, {Novak}, {Freimuth},
  {Mokrousov}, {Rushforth} et~al.}}]{2015arXiv150303765W}
\bibinfo{author}{\bibfnamefont{P.}~\bibnamefont{{Wadley}}},
  \bibinfo{author}{\bibfnamefont{B.}~\bibnamefont{{Howells}}},
  \bibinfo{author}{\bibfnamefont{J.}~\bibnamefont{{Zelezny}}},
  \bibinfo{author}{\bibfnamefont{C.}~\bibnamefont{{Andrews}}},
  \bibinfo{author}{\bibfnamefont{V.}~\bibnamefont{{Hills}}},
  \bibinfo{author}{\bibfnamefont{R.~P.} \bibnamefont{{Campion}}},
  \bibinfo{author}{\bibfnamefont{V.}~\bibnamefont{{Novak}}},
  \bibinfo{author}{\bibfnamefont{F.}~\bibnamefont{{Freimuth}}},
  \bibinfo{author}{\bibfnamefont{Y.}~\bibnamefont{{Mokrousov}}},
  \bibinfo{author}{\bibfnamefont{A.~W.} \bibnamefont{{Rushforth}}},
  \bibnamefont{et~al.}, \bibinfo{journal}{ArXiv e-prints}
  (\bibinfo{year}{2015}), \eprint{1503.03765}.

\bibitem[{\citenamefont{Bode et~al.}(2002)\citenamefont{Bode, Heinze, Kubetzka,
  Pietzsch, Nie, Bihlmayer, Bl\"ugel, and
  Wiesendanger}}]{PhysRevLett.89.237205}
\bibinfo{author}{\bibfnamefont{M.}~\bibnamefont{Bode}},
  \bibinfo{author}{\bibfnamefont{S.}~\bibnamefont{Heinze}},
  \bibinfo{author}{\bibfnamefont{A.}~\bibnamefont{Kubetzka}},
  \bibinfo{author}{\bibfnamefont{O.}~\bibnamefont{Pietzsch}},
  \bibinfo{author}{\bibfnamefont{X.}~\bibnamefont{Nie}},
  \bibinfo{author}{\bibfnamefont{G.}~\bibnamefont{Bihlmayer}},
  \bibinfo{author}{\bibfnamefont{S.}~\bibnamefont{Bl\"ugel}}, \bibnamefont{and}
  \bibinfo{author}{\bibfnamefont{R.}~\bibnamefont{Wiesendanger}},
  \bibinfo{journal}{Phys. Rev. Lett.} \textbf{\bibinfo{volume}{89}},
  \bibinfo{pages}{237205} (\bibinfo{year}{2002}).

\bibitem[{\citenamefont{Gould et~al.}(2004)\citenamefont{Gould, R\"uster,
  Jungwirth, Girgis, Schott, Giraud, Brunner, Schmidt, and
  Molenkamp}}]{PhysRevLett.93.117203}
\bibinfo{author}{\bibfnamefont{C.}~\bibnamefont{Gould}},
  \bibinfo{author}{\bibfnamefont{C.}~\bibnamefont{R\"uster}},
  \bibinfo{author}{\bibfnamefont{T.}~\bibnamefont{Jungwirth}},
  \bibinfo{author}{\bibfnamefont{E.}~\bibnamefont{Girgis}},
  \bibinfo{author}{\bibfnamefont{G.~M.} \bibnamefont{Schott}},
  \bibinfo{author}{\bibfnamefont{R.}~\bibnamefont{Giraud}},
  \bibinfo{author}{\bibfnamefont{K.}~\bibnamefont{Brunner}},
  \bibinfo{author}{\bibfnamefont{G.}~\bibnamefont{Schmidt}}, \bibnamefont{and}
  \bibinfo{author}{\bibfnamefont{L.~W.} \bibnamefont{Molenkamp}},
  \bibinfo{journal}{Phys. Rev. Lett.} \textbf{\bibinfo{volume}{93}},
  \bibinfo{pages}{117203} (\bibinfo{year}{2004}).

\bibitem[{\citenamefont{Velev et~al.}(2005)\citenamefont{Velev, Sabirianov,
  Jaswal, and Tsymbal}}]{PhysRevLett.94.127203}
\bibinfo{author}{\bibfnamefont{J.}~\bibnamefont{Velev}},
  \bibinfo{author}{\bibfnamefont{R.~F.} \bibnamefont{Sabirianov}},
  \bibinfo{author}{\bibfnamefont{S.~S.} \bibnamefont{Jaswal}},
  \bibnamefont{and} \bibinfo{author}{\bibfnamefont{E.~Y.}
  \bibnamefont{Tsymbal}}, \bibinfo{journal}{Phys. Rev. Lett.}
  \textbf{\bibinfo{volume}{94}}, \bibinfo{pages}{127203}
  (\bibinfo{year}{2005}).

\bibitem[{\citenamefont{Sokolov et~al.}(2006)\citenamefont{Sokolov, Zhang,
  Tsymbal, Redepenning, and Doudin}}]{Sokolov}
\bibinfo{author}{\bibfnamefont{A.}~\bibnamefont{Sokolov}},
  \bibinfo{author}{\bibfnamefont{C.}~\bibnamefont{Zhang}},
  \bibinfo{author}{\bibfnamefont{E.~Y.} \bibnamefont{Tsymbal}},
  \bibinfo{author}{\bibfnamefont{J.}~\bibnamefont{Redepenning}},
  \bibnamefont{and} \bibinfo{author}{\bibfnamefont{B.}~\bibnamefont{Doudin}},
  \bibinfo{journal}{Nat Nano} \textbf{\bibinfo{volume}{2}},
  \bibinfo{pages}{171} (\bibinfo{year}{2006}).

\bibitem[{\citenamefont{Iancu et~al.}(2014)\citenamefont{Iancu, Braun,
  Schouteden, and Van~Haesendonck}}]{PhysRevLett.113.106102}
\bibinfo{author}{\bibfnamefont{V.}~\bibnamefont{Iancu}},
  \bibinfo{author}{\bibfnamefont{K.-F.} \bibnamefont{Braun}},
  \bibinfo{author}{\bibfnamefont{K.}~\bibnamefont{Schouteden}},
  \bibnamefont{and}
  \bibinfo{author}{\bibfnamefont{C.}~\bibnamefont{Van~Haesendonck}},
  \bibinfo{journal}{Phys. Rev. Lett.} \textbf{\bibinfo{volume}{113}},
  \bibinfo{pages}{106102} (\bibinfo{year}{2014}).

\bibitem[{\citenamefont{Lei et~al.}(2013)\citenamefont{Lei, Feng, Li, Li, Zhao,
  Wang, Yang, and Hou}}]{rectifier}
\bibinfo{author}{\bibfnamefont{S.}~\bibnamefont{Lei}},
  \bibinfo{author}{\bibfnamefont{W.}~\bibnamefont{Feng}},
  \bibinfo{author}{\bibfnamefont{B.}~\bibnamefont{Li}},
  \bibinfo{author}{\bibfnamefont{Q.}~\bibnamefont{Li}},
  \bibinfo{author}{\bibfnamefont{A.}~\bibnamefont{Zhao}},
  \bibinfo{author}{\bibfnamefont{B.}~\bibnamefont{Wang}},
  \bibinfo{author}{\bibfnamefont{J.}~\bibnamefont{Yang}}, \bibnamefont{and}
  \bibinfo{author}{\bibfnamefont{J.~G.} \bibnamefont{Hou}},
  \bibinfo{journal}{Appl. Phys. Lett.} \textbf{\bibinfo{volume}{102}},
  \bibinfo{eid}{163506} (\bibinfo{year}{2013}).

\bibitem[{\citenamefont{Maslyuk et~al.}(2006)\citenamefont{Maslyuk, Bagrets,
  Meded, Arnold, Evers, Brandbyge, Bredow, and Mertig}}]{PhysRevLett.97.097201}
\bibinfo{author}{\bibfnamefont{V.~V.} \bibnamefont{Maslyuk}},
  \bibinfo{author}{\bibfnamefont{A.}~\bibnamefont{Bagrets}},
  \bibinfo{author}{\bibfnamefont{V.}~\bibnamefont{Meded}},
  \bibinfo{author}{\bibfnamefont{A.}~\bibnamefont{Arnold}},
  \bibinfo{author}{\bibfnamefont{F.}~\bibnamefont{Evers}},
  \bibinfo{author}{\bibfnamefont{M.}~\bibnamefont{Brandbyge}},
  \bibinfo{author}{\bibfnamefont{T.}~\bibnamefont{Bredow}}, \bibnamefont{and}
  \bibinfo{author}{\bibfnamefont{I.}~\bibnamefont{Mertig}},
  \bibinfo{journal}{Phys. Rev. Lett.} \textbf{\bibinfo{volume}{97}},
  \bibinfo{pages}{097201} (\bibinfo{year}{2006}).

\bibitem[{\citenamefont{Gr\"unewald et~al.}(2011)\citenamefont{Gr\"unewald,
  Wahler, Schumann, Michelfeit, Gould, Schmidt, W\"urthner, Schmidt, and
  Molenkamp}}]{PhysRevB.84.125208}
\bibinfo{author}{\bibfnamefont{M.}~\bibnamefont{Gr\"unewald}},
  \bibinfo{author}{\bibfnamefont{M.}~\bibnamefont{Wahler}},
  \bibinfo{author}{\bibfnamefont{F.}~\bibnamefont{Schumann}},
  \bibinfo{author}{\bibfnamefont{M.}~\bibnamefont{Michelfeit}},
  \bibinfo{author}{\bibfnamefont{C.}~\bibnamefont{Gould}},
  \bibinfo{author}{\bibfnamefont{R.}~\bibnamefont{Schmidt}},
  \bibinfo{author}{\bibfnamefont{F.}~\bibnamefont{W\"urthner}},
  \bibinfo{author}{\bibfnamefont{G.}~\bibnamefont{Schmidt}}, \bibnamefont{and}
  \bibinfo{author}{\bibfnamefont{L.~W.} \bibnamefont{Molenkamp}},
  \bibinfo{journal}{Phys. Rev. B} \textbf{\bibinfo{volume}{84}},
  \bibinfo{pages}{125208} (\bibinfo{year}{2011}).

\bibitem[{\citenamefont{Li et~al.}(2015)\citenamefont{Li, Bai, Chen, Zhou, Shi,
  Zhang, Ding, Hou, Schwarzacher, Nichols et~al.}}]{Jijun}
\bibinfo{author}{\bibfnamefont{J.-J.} \bibnamefont{Li}},
  \bibinfo{author}{\bibfnamefont{M.-L.} \bibnamefont{Bai}},
  \bibinfo{author}{\bibfnamefont{Z.-B.} \bibnamefont{Chen}},
  \bibinfo{author}{\bibfnamefont{X.-S.} \bibnamefont{Zhou}},
  \bibinfo{author}{\bibfnamefont{Z.}~\bibnamefont{Shi}},
  \bibinfo{author}{\bibfnamefont{M.}~\bibnamefont{Zhang}},
  \bibinfo{author}{\bibfnamefont{S.-Y.} \bibnamefont{Ding}},
  \bibinfo{author}{\bibfnamefont{S.-M.} \bibnamefont{Hou}},
  \bibinfo{author}{\bibfnamefont{W.}~\bibnamefont{Schwarzacher}},
  \bibinfo{author}{\bibfnamefont{R.~J.} \bibnamefont{Nichols}},
  \bibnamefont{et~al.}, \bibinfo{journal}{J. Am. Chem. Soc.}
  \textbf{\bibinfo{volume}{137}}, \bibinfo{pages}{5923} (\bibinfo{year}{2015}).

\bibitem[{\citenamefont{Sun et~al.}(2014)\citenamefont{Sun, Ehrenfreund, and
  Valy~Vardeny}}]{C3CC47126H}
\bibinfo{author}{\bibfnamefont{D.}~\bibnamefont{Sun}},
  \bibinfo{author}{\bibfnamefont{E.}~\bibnamefont{Ehrenfreund}},
  \bibnamefont{and}
  \bibinfo{author}{\bibfnamefont{Z.}~\bibnamefont{Valy~Vardeny}},
  \bibinfo{journal}{Chem. Commun.} \textbf{\bibinfo{volume}{50}},
  \bibinfo{pages}{1781} (\bibinfo{year}{2014}).

\bibitem[{\citenamefont{Wang et~al.}(2005)\citenamefont{Wang, Acioli, and
  Jellinek}}]{541378242}
\bibinfo{author}{\bibfnamefont{J.}~\bibnamefont{Wang}},
  \bibinfo{author}{\bibfnamefont{P.~H.} \bibnamefont{Acioli}},
  \bibnamefont{and} \bibinfo{author}{\bibfnamefont{J.}~\bibnamefont{Jellinek}},
  \bibinfo{journal}{J. Am. Chem. Soc.} \textbf{\bibinfo{volume}{127}},
  \bibinfo{pages}{2812} (\bibinfo{year}{2005}).

\bibitem[{\citenamefont{Mokrousov et~al.}(2007)\citenamefont{Mokrousov,
  Atodiresei, Bihlmayer, Heinze, and Blügel}}]{0957-4484-18-49-495402}
\bibinfo{author}{\bibfnamefont{Y.}~\bibnamefont{Mokrousov}},
  \bibinfo{author}{\bibfnamefont{N.}~\bibnamefont{Atodiresei}},
  \bibinfo{author}{\bibfnamefont{G.}~\bibnamefont{Bihlmayer}},
  \bibinfo{author}{\bibfnamefont{S.}~\bibnamefont{Heinze}}, \bibnamefont{and}
  \bibinfo{author}{\bibfnamefont{S.}~\bibnamefont{Blügel}},
  \bibinfo{journal}{Nanotechnology} \textbf{\bibinfo{volume}{18}},
  \bibinfo{pages}{495402} (\bibinfo{year}{2007}).

\bibitem[{\citenamefont{Hardrat
  et~al.}(2012{\natexlab{a}})\citenamefont{Hardrat, Wang, Freimuth, Mokrousov,
  and Heinze}}]{PhysRevB.85.245412}
\bibinfo{author}{\bibfnamefont{B.}~\bibnamefont{Hardrat}},
  \bibinfo{author}{\bibfnamefont{N.-P.} \bibnamefont{Wang}},
  \bibinfo{author}{\bibfnamefont{F.}~\bibnamefont{Freimuth}},
  \bibinfo{author}{\bibfnamefont{Y.}~\bibnamefont{Mokrousov}},
  \bibnamefont{and} \bibinfo{author}{\bibfnamefont{S.}~\bibnamefont{Heinze}},
  \bibinfo{journal}{Phys. Rev. B} \textbf{\bibinfo{volume}{85}},
  \bibinfo{pages}{245412} (\bibinfo{year}{2012}{\natexlab{a}}).

\bibitem[{\citenamefont{Hardrat
  et~al.}(2012{\natexlab{b}})\citenamefont{Hardrat, Freimuth, Heinze, and
  Mokrousov}}]{PhysRevB.86.165449}
\bibinfo{author}{\bibfnamefont{B.}~\bibnamefont{Hardrat}},
  \bibinfo{author}{\bibfnamefont{F.}~\bibnamefont{Freimuth}},
  \bibinfo{author}{\bibfnamefont{S.}~\bibnamefont{Heinze}}, \bibnamefont{and}
  \bibinfo{author}{\bibfnamefont{Y.}~\bibnamefont{Mokrousov}},
  \bibinfo{journal}{Phys. Rev. B} \textbf{\bibinfo{volume}{86}},
  \bibinfo{pages}{165449} (\bibinfo{year}{2012}{\natexlab{b}}).

\bibitem[{\citenamefont{Vosko et~al.}(1980)\citenamefont{Vosko, Wilk, and
  Nusair}}]{doi:10.1139/p80-159}
\bibinfo{author}{\bibfnamefont{S.~H.} \bibnamefont{Vosko}},
  \bibinfo{author}{\bibfnamefont{L.}~\bibnamefont{Wilk}}, \bibnamefont{and}
  \bibinfo{author}{\bibfnamefont{M.}~\bibnamefont{Nusair}},
  \bibinfo{journal}{Can. J. Phys.} \textbf{\bibinfo{volume}{58}},
  \bibinfo{pages}{1200} (\bibinfo{year}{1980}).

\bibitem[{\citenamefont{Mokrousov et~al.}(2005)\citenamefont{Mokrousov,
  Bihlmayer, and Bl\"ugel}}]{PhysRevB.72.045402}
\bibinfo{author}{\bibfnamefont{Y.}~\bibnamefont{Mokrousov}},
  \bibinfo{author}{\bibfnamefont{G.}~\bibnamefont{Bihlmayer}},
  \bibnamefont{and} \bibinfo{author}{\bibfnamefont{S.}~\bibnamefont{Bl\"ugel}},
  \bibinfo{journal}{Phys. Rev. B} \textbf{\bibinfo{volume}{72}},
  \bibinfo{pages}{045402} (\bibinfo{year}{2005}).

\bibitem[{\citenamefont{Abate and Asdente}(1965)}]{PhysRev.140.A1303}
\bibinfo{author}{\bibfnamefont{E.}~\bibnamefont{Abate}} \bibnamefont{and}
  \bibinfo{author}{\bibfnamefont{M.}~\bibnamefont{Asdente}},
  \bibinfo{journal}{Phys. Rev.} \textbf{\bibinfo{volume}{140}},
  \bibinfo{pages}{A1303} (\bibinfo{year}{1965}).

\end{thebibliography}
\end{document}